\documentclass[12pt]{iopart}

\usepackage{iopams}
\usepackage{graphicx}
\begin{document}
\title[]{Controlling single diamond NV color center photoluminescence spectrum with a Fabry-Perot microcavity}

\author{Yannick Dumeige$^1$, Romain All\'eaume$^2$, Philippe~Grangier$^3$, Fran\c cois Treussart$^4$ and Jean-Fran\c cois Roch$^4$}

\address{$^1$Universit\'e Europ\'eenne de Bretagne, Laboratoire Foton, CNRS UMR 6082 Foton, Enssat, 6 rue de Kerampont, 22305 Lannion Cedex, France}
\address{$^2$ Laboratoire Traitement et Communication de l'Information, CNRS UMR 5141, Institut T\'el\'ecom / T\'el\'ecom ParisTech, 46 rue Barrault, 75634 Paris Cedex, France}
\address{$^3$Laboratoire Charles Fabry de l'Institut d'Optique, CNRS UMR 8501, Institut d'Optique Graduate School, Campus Polytechnique - RD 128, 2 avenue Augustin Fresnel 91127 Palaiseau Cedex, France}
\address{$^4$Laboratoire de Photonique Quantique et Mol\'eculaire, CNRS UMR 8537, \'Ecole Normale Sup\'erieure de Cachan, 61 avenue du Pr\'esident Wilson, 94235 Cachan Cedex, France}
\ead{yannick.dumeige@univ-rennes1.fr}

\begin{abstract}
We present both theoretical and experimental results on fluorescence of single defect centers in diamond nanocrystals embedded in a planar dielectric microcavity. From a theoretical point of view, we show that the overall fluorescence collection efficiency using moderate numerical aperture microscope objective can be enhanced by using a low quality factor microcavity. This could be used in particular for low temperature applications where the numerical aperture of collection microscope objectives is limited due to the experimental constraints. We experimentally investigate the control of the fluorescence spectrum of the emitted light from a single center. We show the simultaneous narrowing of the room temperature broadband emission spectrum and the increase of the fluorescence spectral density. 
\end{abstract}

\maketitle

\section{Introduction}

Single nitrogen-vacancy (NV) defect centers in diamond is a promising candidate for solid state quantum computing and quantum information processing (QIP) applications~\cite{OBrien10}. It has been shown that NV center can be used as a reliable source of single photons~\cite{Kurtsiefer00} for quantum key distribution  (QKD)~\cite{Beveratos02, Alleaume04} and single photon interferences~\cite{Jacques07}. 
More recently, its potentialities as a spin qubit with a very long spin coherence time~\cite{Kennedy03} have been successfully demonstrated in the context of the coherent coupling of NV$^-$ electronic spin to single neighbouring nuclear spins~\cite{Jelezko04,Lukin06}. It takes advantage of both the spin coupling to optical transitions and the optical readout of the associated electron spin resonance~\cite{Jelezko04bis}.
The spin and optical properties of a single NV center were also used to probe magnetic fields at the nanoscale~\cite{Balasubramanian08,Maze08}.

In both QIP or QKD applications, the efficiency of the collection of the radiated photons is crucial. The rate of the qubit exchange in QKD or the efficiency of optical read-out in QIP strongly depends on the external quantum yield. To control their spontaneous emission rate or to increase their fluorescence collection, solid state individual emitters can be embedded in an optical microcavity~\cite{Moreau01, Barnes02}. For NV center systems, different configurations of coupling with a microcavity have been investigated. Some efforts have been made in designing diamond photonic crystals microcavity~\cite{Hanic06, Kreuzer08, Zhang09}. Microcavity with a quality factors $Q$ around 600 have been fabricated in a thin photonic crystals membrane~\cite{Wang07PC}. Due to the difficulty to obtain monolithic microcavities from bulk or nanocrystalline diamond~\cite{Wang07} others approaches based on hybrid structures have been used. 
Normal mode splitting has been observed in a system consisting of diamond nanocrystals coupled to the whispering gallery mode (WGM) of a silica  microsphere~\cite{Park06}. The coupling of a single NV center to a high-$Q$ factor WGM resonator has been observed in silica microdisk~\cite{Fu08} or microsphere~\cite{Schietinger09, Schietinger09b}. By using a fiber taper, a single diamond nanocrystal has been positioned onto a silica WGM toroid which demonstrates the possibility to manipulate non-classical emitters in order to couple them with high-$Q$ microresonators~\cite{Gregor09}. GaP WGM microdisk have also been used to efficiently collect at the chip scale the fluorescence of NV centers~\cite{Barclay09}. To reach very low mode volume, an optimization scheme of $\mathrm{TiO_2/SiO_2}$ micropillar bragg cavity has been proposed~\cite{Zhang09b}. Alternatively, hybrid structure involving photonic crystals in Silicon Nitride~\cite{McCutcheon08} or GaP~\cite{Barclay09b} enabling coupling to NV centers have been proposed to reach high $Q$-factors. In this context the controlled coupling of a single diamond nanocrystal to a planar $\mathrm{Si_3N_4}$ photonic crystal heterostructure~\cite{Barth09} and the emission spectrum modification of a single NV center embedded in an opal photonic crystal structure made of polystyrene microsphere~\cite{Stewart09} were experimentally demonstrated. Finally one-dimensional photonic nanostructures consisting of diamond nanowires can also be used to enhance the single photon emission from NV centers~\cite{Babinec10}. 

In most of these previous cases, the goals were to obtain simultaneously a high-Q factors and a small mode volume in order to enhance the Purcell factor~\cite{Purcell46} and reach the regime in which the emitter spontaneous emission rate is strongly affected by its coupling to a cavity mode. This results in enhanced spectral properties of the single-photon source well adapted to GHz repetition rate QKD~\cite{Su08}.

However, the emission rate can be not only strongly modified but also the radiation pattern of the emitter~\cite{Drexhage70, Lukosz79}. The thorough investigation of the modification of the fluorescence spectrum and the radiative pattern by photonic structures is more recent, in particular for broadband emission. In that later case, the dispersion associated to its dielectric environment leads to a renormalization of the emitter spectrum, which is such that a simple planar microcavity can strongly modify the emission spectrum of an organic dye~\cite{Enderlein02, Steiner05, Steiner08, Chizhik09} or of a colloidal CdSe/ZnS nanocrystal~\cite{Qualtieri09}.\\

In this work we show that the use of a planar microcavity lead to an enhancement of the collection efficiency of the single NV center photoluminescence in practical configuration of QKD or QIP. We also demonstrate that the microcavity can be used to tune the emission spectrum. The article consists of two theoretical and experimental parts. The first part begins with the description of the planar microcavity structure. We then present the calculation method used to determine the fluorescence power collected by the microscope objective from a single NV color center located either on top of a high-reflectivity mirror or inside the microcavity. We consider both cases of the monochromatic emission (standing for cryogenic operation of the NV center emission) and the room-temperature broadband emission.
In the second part, we describe the experimental setup and the observations showing the modifications of the emission associated to the coupling of a single NV color center in a nanodiamond to a single longitudinal-mode of the planar dielectric microcavity.

\section{Calculation of the total fluorescence power $P_{\rm tot}$ collected from a single emitting dipole coupled to a planar Fabry-Perot microcavity}

\label{sec:calculation}

The collection of light from a single emitter embedded in a dielectric material with a high index of refraction, as NV color center in bulk diamond or semiconductor quantum dots, shares similar issues than those of semiconductor light emitting diodes (LED). 
They mainly come from the light trapping due to total internal reflections. A lot of theoretical and experimental studies have shown that light collection efficiency of LED can be enhanced by using a planar microcavity~\cite{Benisty98a, Benisty98b}. This concept has been adapted to high index of refraction solid-state single-photon sources~\cite{Barnes02}. 

In the case of NV color center in diamond nanocrystals it has been pointed out that refraction is irrelevant due to the small dimensions of the nanocrystals compared to the emission wavelength and that the emitter can be assumed as a point source emitting in the surrounding dielectric medium~\cite{Beveratos01}. This feature enhances the collection efficiency. 

In this section we will show that the use of a resonant microcavity combined with the nanocrystal approach can also increase the collection efficiency for NV defect centers. By using a high reflectivity rear mirror the isotropic radiation is redirected in the same half space~\cite{Begon00}, while the microcavity simultaneously narrows the emission spectrum~\cite{Rigneault96}, both effects resulting in a larger spectral density of the collected light intensity . 
Finally for an appropriate choice of the emitting dipole position and orientation, one can also benefit from the local electromagnetic field enhancement~\cite{Benisty98a}.

\subsection{Microcavity~--~single-emitter structure}

\begin{figure}[htbp]
\centering\includegraphics[width=13cm]{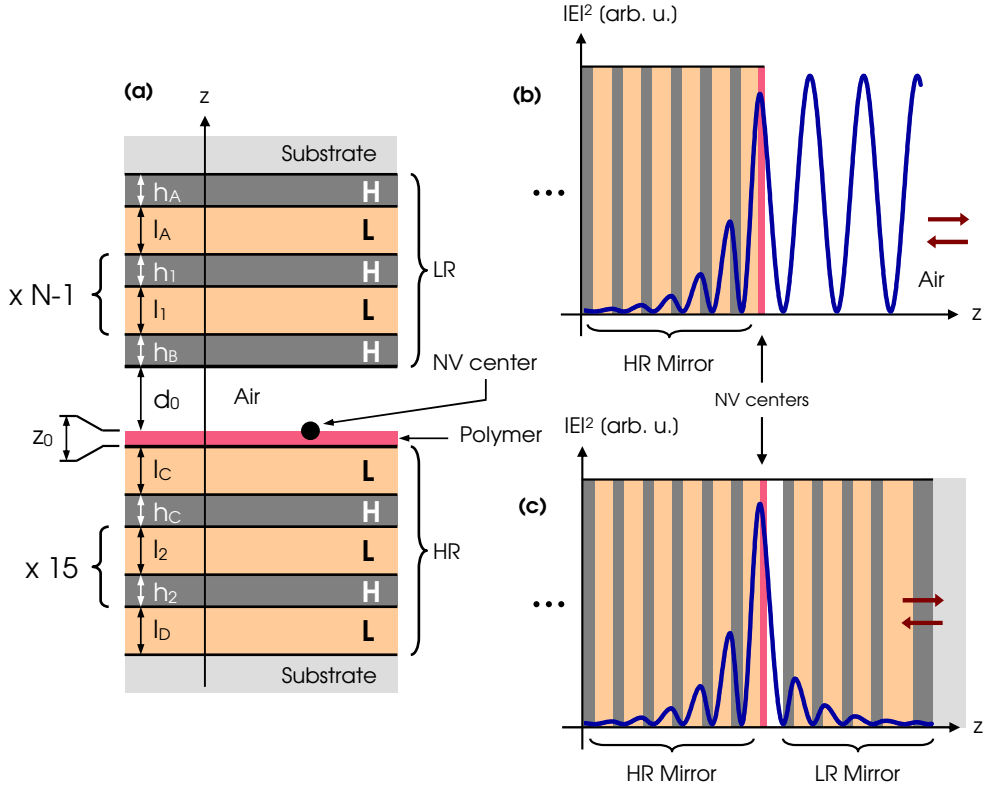}
\caption{(a) Scheme of the single-photon source microcavity consisting of a diamond nanocrystal containing a single NV center, positioned in the air gap between the two mirrors of a planar Fabry-Perot microcavity. 
The lower high reflectivity (HR) Bragg mirror is composed of a fixed number of 16 periods, whereas the upper low reflectivity (LR) mirror is made of $N$ periods. Referring to the figure annotations, the thicknesses values of the different layers are: $h_{\rm A}=134~\mathrm{nm}$, $l_{\rm A}=171.7~\mathrm{nm}$, $h_1=h_2=h_{\rm B}=73.1~\mathrm{nm}$, $l_1=l_2=113.7~\mathrm{nm}$, $h_{\rm C}=75.3~\mathrm{nm}$, $l_{\rm C}=110.1~\mathrm{nm}$ and $l_{\rm D}=117.1~\mathrm{nm}$. 
The single NV color center is assumed to be located on the top of the polymer layer of thickness $z_0=25~\mathrm{nm}$. $d_0$ is the adjustable air gap thickness. The low index of refraction ($n_{\rm \ell}=1.4745$ at $\lambda=680$~nm) layers are marked L and the high ($n_{\rm h}=2.2925$) index layers H. The refractive indices of the polymer and of the substrate are respectively $n_{\rm p}=1.49$ and $n_{\rm s}=1.4556$. We also give the optical intensity distribution $\left|E(z)\right|^2$ at $\lambda_{\mathrm{NV}}^{\mathrm{max}}$ for a plave wave excitation in normal incidence. The calculation have been made using the transfer-matrix method \cite{Yeh77} for (a) the HR mirror and (b) the microcavity with $\lambda_0=\lambda_{\mathrm{NV}}^{\mathrm{max}}$ and $N=4$.}\label{structure}
\end{figure}

The single-photon source described in figure \ref{structure} consists of a nanocrystal embedded in a thin polymer layer deposited on a dielectric high reflectivity (HR) Bragg mirror made of 16 periods of two materials of high $n_{\rm h}$ and low $n_{\rm \ell}$ indices of refraction~\cite{Beveratos02b}.

The cavity is obtained using a second low reflectivity (LR) mirror made of $N<16$ periods of the same materials. The resonant wavelength of the microcavity can be tuned by changing the air gap $d_0$ between the LR Bragg mirror and the HR mirror. The thicknesses of the different layers of the Bragg mirrors have been chosen to obtain a central wavelength $\lambda_0$ around $\lambda_{\rm NV}^{\rm max}=680~\mathrm{nm}$ corresponding to the peak emission of the NV$^-$ center at room temperature. The structure has also been optimized to maximize the electric field amplitude in the polymer layer containing the nanocrystals. We give in figure \ref{structure} the electric field amplitude distribution for a normal incidence plane wave in the two cases we extensively study in this paper: (a) HR mirror without second mirror (b) microcavity with $N=4$. 

\begin{figure}[htbp]
\centering\includegraphics[width=11cm]{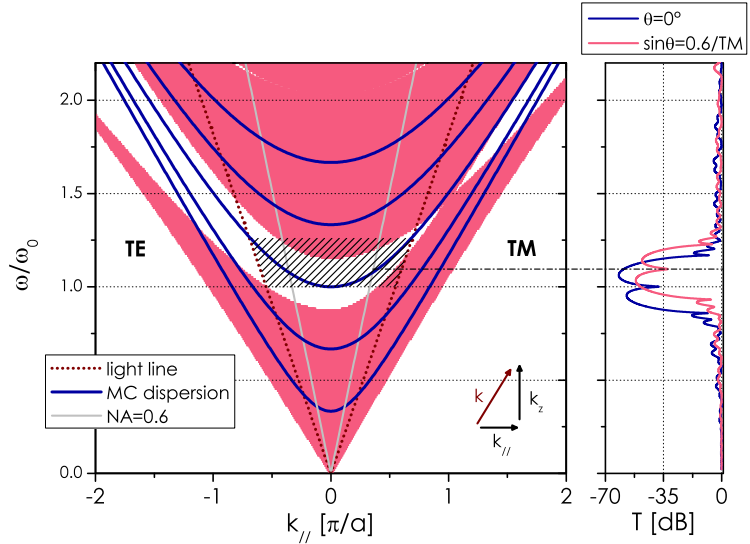}
\caption{Graphical representation of the Bragg mirror and microcavity dispersion relations (left) and of the transmission spectrum (right). TE (resp. TM) polarization case is displayed on the negative (resp. positive) $k_\parallel$ side. $a=l_1+h_1$ is the period of the Bragg mirror. The propagation is forbidden in the white areas and allowed in the pink areas. The planar microcavity (MC) dispersion curves are also represented (in blue) assuming an interference order $m_{\rm c}=3$ for $\lambda_0$ in normal incidence. We have added two light lines: the one associated to a full half-space collection (numerical aperture N.A.=1, red dot line) and the other for a restricted collection of light with N.A.=0.6 optics (grey line). 
{\bf Right graph}: transmission spectrum $T(\omega,\theta,\mathrm{TM})$ of the microcavity ($N=4$ case) in $\mathrm{dB}$ units at either normal incidence ($\theta=0^\circ$, blue curve) or in the case of inclined incidence with $\sin\theta=0.6$ and TM polarization (red curve). These curves display dips at their resonant wavelengths.
The dashed region on the left graph indicates the range of frequencies allowed in the case of N.A.=1. Note that for N.A.=0.6, only the frequencies shorter than the resonance frequency are allowed, which explains why the experimental spectra displayed on figure~\ref{spectra} and figure~\ref{experiment} are asymmetric, with a steep falling edge on the high frequency side.
}\label{dispersion}
\end{figure}

Figure~\ref{dispersion} represents the dispersion relation $\omega(k_{\parallel})$ of the Bragg mirrors where $\omega$ is the angular frequency and $k_{\parallel}$ is the in-plane component of the wavector $\bf k$~\cite{Yeh77}. Due to the field penetration in the dielectric mirror the minimal interference order for the planar microcavity is given by $m_{\rm c}=\lfloor \overline{n}/\Delta n \rfloor+1$ with $\Delta n=n_{\rm h}-n_{\rm \ell}$, $\overline{n}=n_{\rm h}n_{\rm \ell}/n_{\rm a}$ and $n_{\rm a}=(n_{\rm \ell}l_1+n_{\rm h}h_1)/(l_1+h_1)$ where $\lfloor x \rfloor$ is the floor function of $x$ \cite{Benisty98a}. With our refractive index values we obtain $m_{\rm c}=3$ which is used to obtain the microcavity dispersion curves displayed in figure~\ref{dispersion} as solid lines. 

The dispersion relation is obtained noting first that $k=\sqrt{k_{\parallel}^2+k_z^2}$. Then, if we call $L_{\rm cav}$ and $n_{\rm cav}$ the effective length and refractive index of the cavity taking into account the penetration depth in the Bragg mirrors, we obtain $2k_zL_{\rm cav}=2m_{\rm c}\pi$ for a resonant mode and thus:

\begin{equation}
\omega=\frac{c}{n_{\rm cav}}\sqrt{k_{\parallel}^2+\left(\frac{m_{\rm c}\pi}{L_{\rm cav}}\right)^2}.
\end{equation}
Neglecting the polymer layer effect, the two quantities $L_{\rm cav}$ and $n_{\rm cav}$ can be estimated by solving the resonance equation (for $d_0\approx135~\mathrm{nm}$):
\begin{equation}
m_{\rm c}\frac{\lambda_0}{2}=1\times d_0+n_{\rm a}(L_{\rm cav}-d_0)=n_{\rm cav}L_{\rm cav}.
\end{equation}

We notice that with the microcavity design, the dispersion curve referring to $m_c=3$ is fully included in the first forbidden band of the Bragg mirror for both TE and TM polarizations and for the maximal numerical aperture considered here (\emph{i.e.} N.A.=1, corresponding to the light line) and thus also for the value N.A.=0.6 used in the experiments. 
This feature is important for room temperature experiments since in this case the fluorescence spectrum of NV centers is broad (full width at half maximum FWHM$\approx75~\mathrm{nm}$) and therefore requires a broadband operation range. 
We also show the transmission spectrum $T(\omega,\theta,\rm TM)$ obtained using the standard transfer matrix method \cite{Yeh77} for a microcavity with $N=4$ both for the normal incidence, and for $\sin\theta=0.6$ TM polarization cases. Note that due to the unbalanced mirror reflectivity, the resonant transmission is weak even though we assumed lossless materials.

This simple linear analysis~\cite{Barnes02} allows two properties of the microcavity to be revealed: i) for the numerical aperture N.A.=0.6 used in our experiment it is possible to efficiently collect light coupled to the resonant mode of the cavity, ii) in the case of a broadband emission, the blue-shift of the cavity resonance explains the spectrum asymmetry experimentally observed (see Section~\ref{exp}).

\subsection{Calculation method of the total collected power $P_{\rm tot}$ radiated by a single NV color center inside the microcavity}

It is well established since the early studies of NV$^-$ color center~\cite{Davies76}, that it has two orthogonal incoherent electric dipole transitions oriented in the plane perpendicular to the [111] crystallographic direction (along the N-V axis), as exemplified in the context of recent studies on polarization selective excitation of this color center~\cite{Awschalom05,Beausoleil07}.

\begin{figure}[htbp]
\centering\includegraphics[width=11cm]{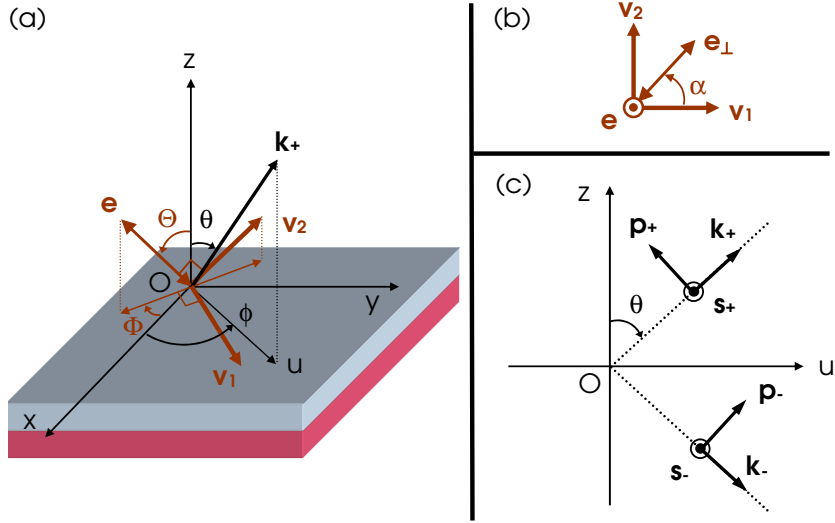}
\caption{(a) O is the position of the orthogonal $(\mathbf{e},\mathbf{e}_{\bot})$ dipoles that we associate to the NV$^-$ color center transition. The orientation of the component $\mathbf{e}$ is described by two angles $(\Theta,\Phi)$ relative to the Bragg mirrors plane $(x,y)$. $\bf k_+$ is the wavevector of the field in the upper half space emitted in the direction $(\theta,\phi)$. (b) $\alpha$ refers to the orientation of $\mathbf{e}_{\bot}$ in the plane $(\mathbf{v}_1,\mathbf{v}_2)$ orthogonal to $\mathbf{e}$. (c) $\bf k_-$ is the wavevector of the field emitted in the same direction but in the lower half space. The unitary vectors are noted $\bf s_{\pm}$ (resp. $\bf p_{\pm}$) for TE (resp. TM) polarization.}
\label{dipole}
\end{figure}

In the following we will model the NV$^-$ center by two identical orthogonal dipoles $(\mathbf{e},\mathbf{e}_{\bot})$. The dipoles are assumed to be located in O. $\mathbf{e}$ orientation is defined by two angles $\Theta$ and $\Phi$ relative to the Bragg mirror plane $(x,y)$, such as $\mathbf{e}=(\sin{\Theta}\cos{\Phi},\sin{\Theta}\sin{\Phi},\cos{\Theta})$ in the sample plane coordinate system as represented in figure \ref{dipole}. Thus the second dipole can be written: $\mathbf{e}_{\bot}=\mathbf{v}_1\cos{\alpha}+\mathbf{v}_2\sin{\alpha}$ with $\mathbf{v}_1=(-\sin{\Phi},\cos{\Phi},0)$ and $\mathbf{v}_2=(-\cos{\Theta}\cos{\Phi},-\cos{\Theta}\sin{\Phi},\sin{\Theta})$. The most favorable configuration to efficiently collect the photoluminescence of the NV$^-$ color center corresponds to $(\Theta=\pi/2,\alpha=0)$ whereas the worst configuration is obtained for $\Theta=0$. 
In the experiment we selected the brightest emitters, which are the one having the strongest in-sample plane dipole components. Therefore, in the following we restrict our calculation of the collected power $P_{\rm tot}$ to this configuration $(\Theta=\pi/2,\alpha=0)$. To calculate the power radiated by the two dipoles in the upper half space ($z>0$), we used the transfer-matrix method of reference~\cite{Benisty98}.

For a given direction of radiation characterized by the wavevector $\mathbf{k_+}$, with coordinate $\mathbf{k_+}=k(\sin{\theta}\cos{\phi},\sin{\theta}\sin{\phi},\cos{\theta})$, the wave can be either TE-polarized along $\mathbf{s_+}$ or TM-polarized along $\mathbf{p_+}$. $\mathbf{s_+}$ is the unit vector orthogonal to the incidence plane $(u,z)$, whereas $\mathbf{p_+}$ is calculated by: 

\begin{equation}\label{def_p}
\mathbf{p_+}=\frac{1}{k}\mathbf{s_+}\times\mathbf{k_+}.
\end{equation}

The wavevector of the field related to $\mathbf{k_+}$ propagating in the opposite direction is noted $\mathbf{k_-}=k(\sin{\theta}\cos{\phi},\sin{\theta}\sin{\phi},-\cos{\theta})$. We also have $\mathbf{s_-}=\mathbf{s_+}$ and the same relation as Eq.~(\ref{def_p}) holds for $\mathbf{p_-}$. According to the method given in reference~\cite{Benisty98} and using the dipole description of references~\cite{Lukosz79} and \cite{Enderlein99} the source terms for the two propagation directions are: 

\begin{equation}\label{sources}
A_+^s=\mathbf{e}\cdot \mathbf{s_+},~A_+^p=\mathbf{e}\cdot \mathbf{p_+},~A_-^s=-\mathbf{e}\cdot \mathbf{s_-}~\mathrm{and}~A_-^p=\mathbf{e}\cdot \mathbf{p_-}.
\end{equation}

We have the same relations for $\mathbf{e}_{\bot}$. Using these expressions for the optical sources it is possible to calculate the spectral density of the radiated power per solid angle $d^3P/(d\Omega d\lambda)$ for a given $\mathbf{k_+}$ by summing the power emitted by the two incoherent dipoles \cite{Benisty98}. Following the method described in reference \cite{Steiner08}, the full radiated power collected by the air microscope objective of numerical aperture N.A.=$\sin{\theta_{\rm M}}$ in a spectral window of full width $\Delta\lambda$ is then evaluated by :

\begin{equation}\label{radiation}
P_{\rm tot}=\int_{\lambda_0-\Delta\lambda/2}^{\lambda_0+\Delta\lambda/2}f(\lambda)d\lambda\int_0^{2\pi}d\phi\int_0^{\theta_{\rm M}}\frac{d^3P}{d\Omega d\lambda}\sin{\theta}d\theta
\end{equation}
where $f(\lambda)$ is the free space emission normalized spectrum function which will be assumed for simplicity to have a Gaussian profile centered at $\lambda_\mathrm{NV}^\mathrm{max}$ with a full width at half maximum $B$.

\subsection{Calculation of $P_{\rm tot}$}\label{Ptot}

Our aim is to show that the light extraction from a single emitter can be increased using a low finesse microcavity in practical cases. Consequently we will take as the reference the case of dipoles located on a single mirror. The advantage of this last configuration with respect to free space is twofold: the mirror redirects the fluorescence in the same half space and if the dipoles are located in a field antinode, their emission rate is increased~\cite{Benisty98a, Begon00}. 
In the following we consider two cases: 1) the monochromatic emission of NV$^-$ at its zero-phonon line (ZPL) wavelength $\lambda_{\rm ZPL}=637$~nm, which corresponds to cryogenic operation, 2) the broadband emission case, associated to room-temperature photoluminescence.

\subsubsection{Monochromatic case}

First we focus on the low temperature configuration. In this particular case, the fluorescence spectrum of the emitter is assumed to be quasi-monochromatic and the ZPL has a bandwidth around $B=0.3~$nm~\cite{Drabenstedt99,Beveratos_these}. 
For simplicity and because our cavity has been optimized for a resonant wavelength $680~\mathrm{nm}$, we have used for the calculation $\lambda_{\rm NV}^{\rm max}$ as the emission wavelength instead of 
$\lambda_{\rm ZPL}$.

\begin{figure}[htbp]
\begin{center}
\includegraphics[width=11cm]{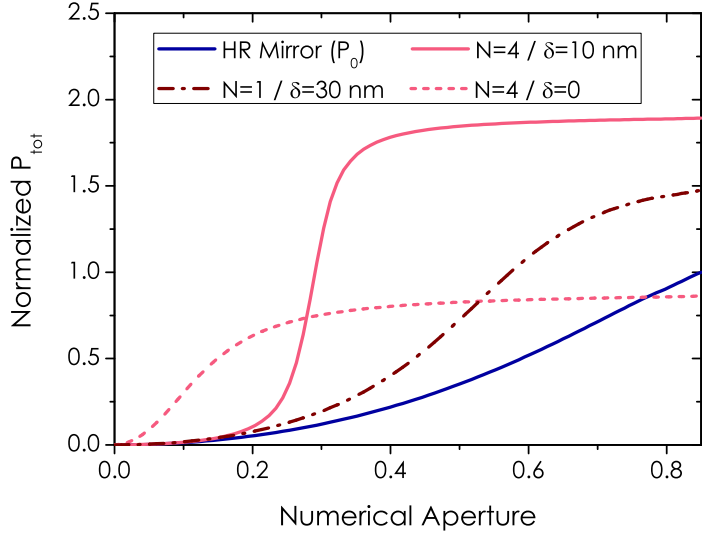}
\caption{Full radiated power $P_{\rm tot}$ in the case of a narrow bandwidth emitter (single NV$^-$ color center emitting at $\lambda_{\rm NV}^{\rm max}$) as a function of the numerical aperture of the collection objective. The calculations have been carried out for a single HR mirror and in the case of the microcavity for two values of $N$ and different values of the microcavity mode -- emitter wavelength detuning $\delta$. The results are normalized to the value of $P_{\rm tot}$ corresponding to the HR mirror case and at the maximal numerical aperture collection (N.A.=0.85).}
\label{froid}
\end{center}
\end{figure}

In order to collect the light across the fused silica substrate of the LR mirror in the case of the microcavity configuration, we use a long working distance objective with a numerical aperture N.A.=0.6. The reference configuration consists of a HR mirror and a microscope objective with N.A.=0.85. We used this low value of N.A. because it is suited to cryogenic setup~\cite{Drabenstedt99}. The simulation results are shown in figure~\ref{froid}. In the case of low finesse cavity ($N=1$) and a detuning $\delta=\lambda_0-\lambda_{\rm NV}^{\rm max}=30$~nm, we observe an increase of about $50~\%$ in the collection efficiency with respect to the reference case. This can be increased by using a better finesse ($\mathcal{F}\approx60$) cavity ($N=4$). Note that in this case, the cavity must be slightly positively detuned ($\delta=10~\mathrm{nm}$) to obtain the maximal collection increase of about $80~\%$. This positive detuning allows the spectral overlap between the emission spectrum and the microcavity dispersion to be optimized as illustrated by the hatched area in figure~\ref{dispersion}. For $\delta=0$ we do not observe any increase in the extracted fluorescence for the maximal numerical aperture.

\subsubsection{Broadband emission}\label{ambiant}

At room temperature, the coupling between the radiative transitions of the NV$^-$ color center and the diamond matrix phonon broadens the emission spectrum yielding a FWHM linewidth of about $B=75~$nm due to the phonon replica.

\begin{figure}[htbp]
\begin{center}
\includegraphics[width= \textwidth]{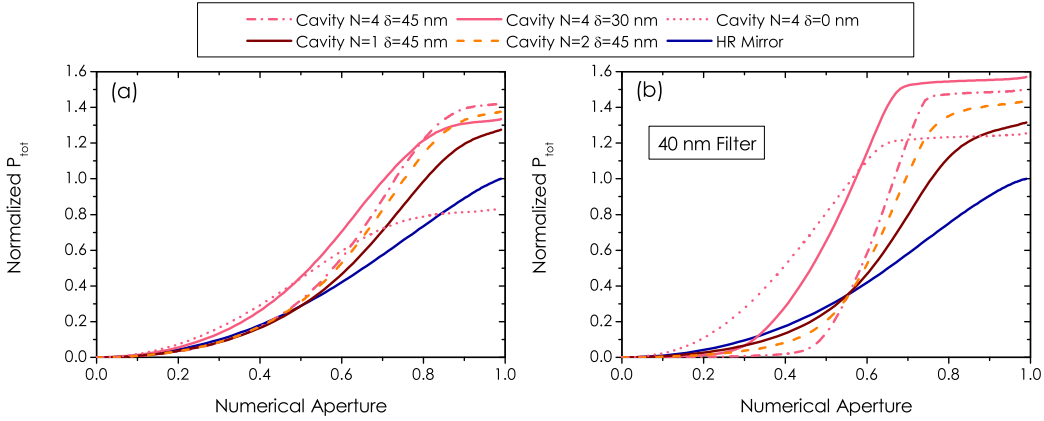}
\caption{Full radiated power $P_{\rm tot}$ in the case of a broadband emitter as a function of the numerical aperture of the collection objective (a) without additional filter (b) with a $\Delta\lambda=40~$nm band-pass filter. The calculations have been carried out for a single HR mirror and for microcavities with three values of $N$ and different values of the microcavity mode -- emitter wavelength detuning $\delta$. The results are normalized for the value corresponding to the HR mirror and the maximal numerical aperture (N.A.=1).}\label{ambiante}
\end{center}
\end{figure}

Figure~\ref{ambiante}.(a) gives the calculation of the full radiated power with no additional filter whereas figure~\ref{ambiante}.(b) represents the same calculations with a $\Delta\lambda=40~\mathrm{nm}$ band-pass filter whose resonant wavelength is chosen in order to increase the overall radiated power for the maximal numerical aperture. This filter could be useful for example in open space QKD applications where it will reject a large amount of background light~\cite{Alleaume04}. As it was the case for monochromatic emission the enhancement is better for $N=4$ and for a positive detuning. The maximal enhancement is weaker than in the previous configuration since in room temperature experiments it is possible to use larger numerical aperture collection objectives with N.A. close to unity. 
Note that the collection efficiency enhancement is improved by the band-pass filter as shown in figure~\ref{ambiante}.(b). {This effect comes from the increase of the coherence length due to the coupling of emitted light to the microcavity mode, similarly to what has already been observed in microcavity LED~\cite{Oulton05}. 
We can underline furthermore this phenomenon by calculating the enhancement ratio $P_{\rm tot}/P_0$ where $P_0$ is the radiated power by the HR mirror at the same N.A. than the one used to calculate $P_{\rm tot}$. The corresponding results are shown in figure~\ref{ambiante1}.(a) for the full spectrum and (b) for the filtered spectrum. We note that the maximal collection efficiency is reached for N.A.$> 0.6$. For N.A.=0.6 and $\delta=30~\mathrm{nm}$ detuning the maximal enhancement is increased by 1.7 for the full bandwidth detection [figure \ref{ambiante1}.(a)] and by 2.75 for the filtered detection [figure \ref{ambiante1}.(b)]. 

At low N.A.$<0.6$ and filtered detection, the collected power is reduced in the microcavity case ($N=4$) with respect to the mirror case for detuning larger than $\delta=30~\mathrm{nm}$. This situation is reversed if one uses smaller detuning values as can be seen for $\delta=30~\mathrm{nm}$ and $\delta=0$ in figure~\ref{ambiante1}.b). 
Indeed for low numerical apertures a large positive detuning has a detrimental role since the cavity resonance would be obtained for too large angles of incidence, out of the collection aperture. 

Finally we have also calculated that for N.A.$<0.82$ and the configuration $N=4$ / $\delta=30~\mathrm{nm}$ the total collected power in the case of the filtered detection is larger than the full bandwidth detection one with only the HR mirror.

\begin{figure}[htbp]
\begin{center}
\includegraphics[width= \textwidth]{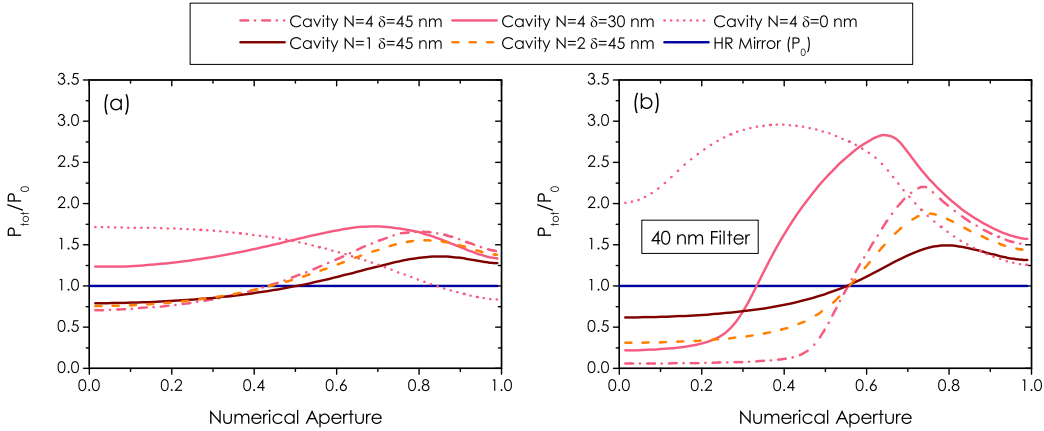}
\caption{Enhancement factor of the total collected power $P_{\rm tot}$ vs. the N.A. in the case of the microcavity with respect to the HR mirror configuration, for which the collected power is $P_0$ (a) without additional filter (b) with a $\Delta\lambda=40~$nm band-pass filter. The power $P_{\rm tot}$ is calculated in the same way than in figure~\ref{ambiante}.}\label{ambiante1}
\end{center}
\end{figure}

Note that the evolution of $P_{\rm tot}$ vs N.A. obtained for the optimal NV color center dipoles orientation holds for all possible orientations. For broadband emission, in the worst configuration $\Theta=0$, and in the case of a numerical aperture $\mathrm{N.A.}=0.6$, the same $P_{\rm tot}$ calculation methods yields a total collected power equal to 53\% of the one of the best configurationin the case of the HR Bragg mirror. Similar results are obtained in the case of a $N=4$ microcavity, for which the total collected power in the worst dipole orientation case corresponds to 50\% of total power in the optimal case.

As a conclusion the simulation shows that the collection of photoluminescence from a single NV$^-$ color center can be increased by its coupling to one low order mode of a planar microcavity. Even in the case of broadband emission the use of a moderate finesse ($\mathcal{F}\approx60$) can enhance the collection efficiency. In addition such coupling also lead to a narrowing of the emitted spectrum and thus to an increase in the coherence length of the emitted light.


\section{Experimental measurement of the spectral properties of the NV color-center coupled to a single mode of the planar microcavity}\label{exp}


\subsection{Experimental setup}

\begin{figure}[htbp]
\centering\includegraphics[width=11cm]{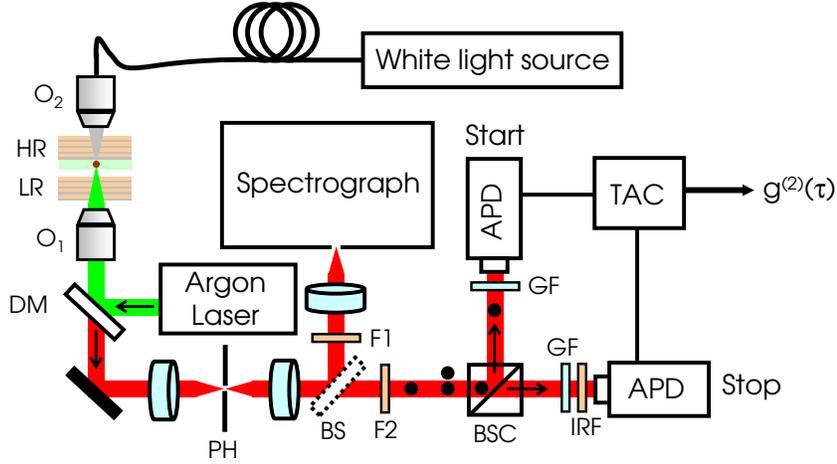}
\caption{Scheme of the setup used to study the fluorescence from a
single NV color center in diamond. $\mathrm{O_1}$: microscope objective
(long working distance, $\times40$; N.A.=0.60); $\mathrm{O_2}$: microscope objective
($\times16$; N.A.=0.32); PH: pinhole of $50~\mathrm{\mu m}$ of diameter; BS: beam splitter plate; F1: glass filter (Schott OG550) transmitting $\lambda>550~\mathrm{nm}$; F2: interference filter transmitting $\lambda>580~\mathrm{nm}$ and removing the remaining light at the pump wavelength; BSC: nonpolarizing 50/50 beam-splitter cube; APDs: Silicon avalanche photodiodes in photon counting regime; IRF: interference filter suppressing near infrared light at $\lambda>740~\mathrm{nm}$, which is combined with two GFs: glass filters (Schott KG4), resulting in the suppression of the optical crosstalk between the APDs due to their breakdown flash~\cite{Weinfurter01}. TAC: Time to Amplitude Converter connected to a multichannel analyser board allowing to extract the time intensity second order correlation function $g^{(2)}$.}\label{setup}
\end{figure}

The single emitter studied in this work is a NV$^-$ color center in type 1b (natural specified content in nitrogen $\approx 200$~ppm) diamond nanocrystals (Micron+ MDA 0-0.1 from De Beers/Element6) with a size smaller than 100~nm, obtain by a high-pressure and high temperature synthesis. In order to increase the natural content of NV center in the nanodiamonds, the particle were irradiated by a high-energy electron beam to generate vacancies and then annealed during 2~hours at $800~^\circ\mathrm{C}$ under vacuum. Then, they were cleaned by oxydative chemicals and dispersed in a propanol solution containing a polymer at low concentration (polyvinylpyrrolidone, $1\%$ w/w). The mean hydrodynamical size of the nanodiamond in the colloidal suspension is 90~nm as measured by dynamic light scattering.
The nanodiamond--polymer solution is finally spincoated onto the HR dielectric mirror forming a $25~\mathrm{nm}$ thick polymer layer with a surface concentration of about one nanocrystal per $10~\mathrm{\mu m}^2$. 
The microcavity is obtained by placing a LR mirror ($N=4$ reflectivity, R$_{LR}$ $\approx$ 90 $\%$) in front of the HR mirror (reflectivity R$_{HR}$ $\approx$ 99.99 $\%$) at a distance that can be adjusted using micrometric screws and piezoelectric actuators. From an experimental of view, the actual values of layer thicknesses for the HR mirror are: $h_2=75.3~\mathrm{nm}$ and $l_2=117.1~\mathrm{nm}$.

Figure~\ref{setup} is a schematic of the confocal optical scanning microscope~\cite{Treussart01} used to study the fluorescence of a single NV center in a dielectric microcavity. The light at $514.5~\mathrm{nm}$ wavelength from a cw argon-ion laser is focused on the nanodiamond-polymer layer through the LR mirror, with a long working distance microscope objective O$_1$ having a N.A.=0.6. The objective $\mathrm{O}_1$ is corrected for the fused silica substrate of the LR mirror. Nevertheless, its optical characteristics can be switched, which allows it to be used without the additional LR mirror. The LR mirror has a reflectivity less than $20\%$ in the range $500-550~\mathrm{nm}$ allowing for efficient excitation at the laser wavelength. The sample is raster scanned with a piezoelectric stage.
The fluorescence from NV centers is collected through the same objective O$_1$, the detection volume being defined by a confocal pinhole. The collected light passes through a dichroic mirror and long-pass filters are used to remove any residual excitation light. The fluorescence light intensity is equally divided into a first beam for spectral analysis by an imaging spectrograph (composed of a grating and a cooled CCD array for spectrum record) and a second beam for measuring the photon statistics using the standard Hanbury-Brown and Twiss (HBT) coincidence setup. The latter measurement allows us to identify the emission of a single color center yielding a perfect photon antibunching. The white light source and the microscope objective O$_2$ are used to measure the microcavity transmission spectrum. Note that in our experimental configuration we obtain a transmission spectrum $\overline{T}(\lambda)$ averaged over the numerical aperture of $\mathrm{O_2}$ and over the two polarizations TE and TM as detailed in \ref{append1}.

\subsection{Narrowing of the spectrum of a single NV$^-$ color center coupled to one microcavity mode}

To identify well isolated fluorescent emitters, we first raster scan the sample. For each fluorescent spot, the uniqueness of the emitter is then checked by the observation of antibunching in the photon statistics of the fluorescent light. Since after the emission of a first photon, it takes a finite time for a single emitter to be excited again and then it spontaneously emits a second photon, the antibunching effect appears as a dip around zero delay $\tau=0$ in the normalized intensity $I(t)$ time autocorrelation function:
\begin{equation}
g^{(2)}(\tau)\equiv\frac{\left\langle I(t)I(t+\tau)\right\rangle}{\left\langle I(t)\right\rangle^2}.
\end{equation}

The HBT detection setup allows one to record the histogram of time delays between two consecutively detected photons on each APD. Considering
our detection efficiency of a few percentages, this histogram is a very good approximation of the intensity autocorrelation function $g^{(2)}(\tau)$.

Figure~\ref{spectra} shows different emission spectra measured for the same single NV center while changing the cavity length. The first spectrum at the bottom is obtained for a closed cavity (i.e., HR and LR mirror into optical contact). In this configuration there is no resonant mode in the Bragg mirror stop-band and the fluorescence of the center is inhibited.

\begin{figure}[htbp]
\begin{center}
\includegraphics[width=11cm]{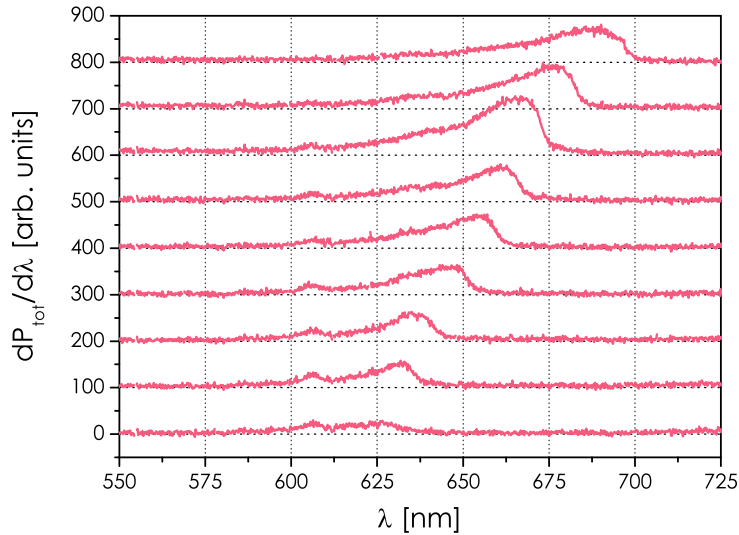}
\caption{Emission spectra of a single NV$^-$ center for different values of the air gap $d_0$ between the LR and HR mirrors. From the bottom of the graph to the top the cavity length is increased using piezoelectric actuators. We start from a configuration where the two mirrors are almost in contact, and pass trough a distance bewteen the mirror yielding a maximum intensity with a narrowing of the emission spectrum compared to free space emission. The small peak at $607~\mathrm{nm}$ is the two-phonons Raman scattering line of the diamond nanocrystal matrix as observed in previous work~\cite{Dumeige04}, corresponding to a Raman shift of 1480~cm$^{-1}$, different to the one of 1335~cm$^{-1}$ in the case of bulk diamond crystal, probably due to the nanocrystalline form.}\label{spectra}
\end{center}
\end{figure}

By opening the cavity we increase the resonant wavelength corresponding to $m_{\rm c}=3$ and we obtain a resonant mode whose wavelength is located inside the emission spectrum of the NV$^-$ center. We observe that the emitted fluorescence intensity is maximal with a noticeable shrinking of the spectrum, when the resonant wavelength $\lambda_0$ is slightly detuned from the free space NV$^-$ center emission maximum $\lambda_{\rm NV}^{\rm max}$ (see third curve starting from top of figure~\ref{spectra}).

Figure \ref{experiment}.(a) shows the detailed analysis of one of the optimal configurations. The experimental transmission $\overline{T}(\lambda)$ has been measured and this optimal case corresponds to the situation where the cavity resonant wavelength is almost tuned at the emission maximum. Here, the optimal detuning $\delta$ is almost zero since the experimental value of N.A. is only 0.6. Figure~\ref{experiment}.(b) displays the cavity emission spectrum and the HR mirror emission spectrum (used as reference) recorded in the same experimental conditions. In particular the saturation of the single NV center has been reached for the two configurations. Finally figures~\ref{experiment}.(c) and \ref{experiment}.(d) are time intensity second order correlation functions $g^{(2)}$ associated to the emission of single NV center, for the mirror and cavity configurations respectively. The vanishing dip at zero delay $\tau=0$ in the $g^{(2)}$ function insures that we address single color center in both cases.

\begin{figure}[htbp]
\begin{center}
\includegraphics[width=13cm]{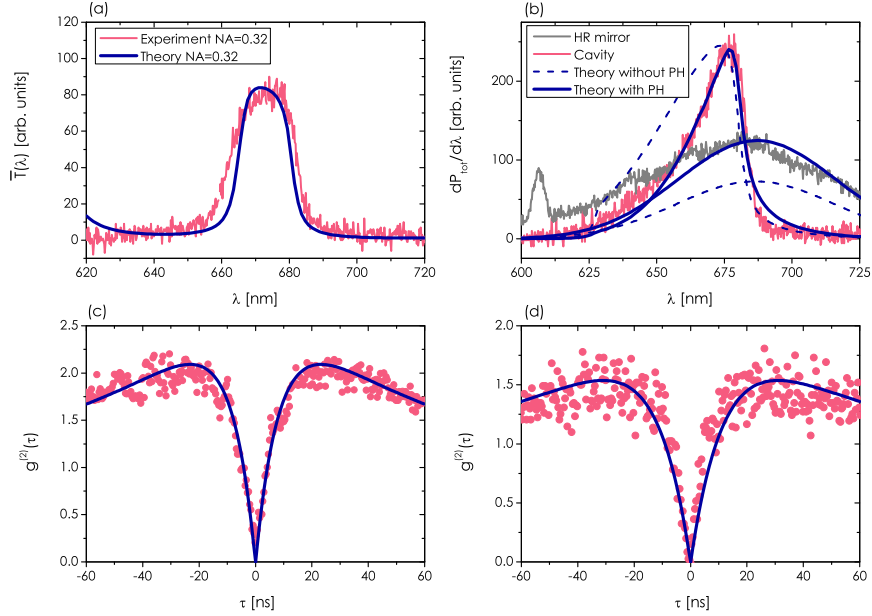}
\caption{(a) Transmission spectrum of the microcavity obtained with its illumination by a white light source at N.A.=0.32 (see figure~\ref{setup}). We also show a calculated transmission spectrum obtained for $d_0=124.1~\mathrm{nm}$. (b) Emission spectra of a single NV centers on a HR mirror (grey plot) or inside a cavity (pink plot). The calculations are normalized using the microcavity spectrum. We give two sets of calculated spectra i) dash curve: without taking into account the pinhole (PH) spatial filtering ii) full line: taking into account the confocal PH filtering and a negative defocusing. The small peak around $607~\mathrm{nm}$ of graph b) is a Raman scattering line of the diamond matrix. Time intensity correlation function $g^{(2)}$ displaying photon antibunching in two cases: (c) single NV center on a HR mirror (d) single NV center inside the microcavity. The fits (blue lines) of the correlation functions have been carried out using the model given in reference~\cite{Beveratos00}.}\label{experiment}
\end{center}
\end{figure}

The fit of the transmission spectrum $\overline{T}(\lambda)$ of the cavity as been obtained only adjusting the value of the air gap thickness (here $d_0=124.1~\mathrm{nm}$ and thus $\delta=1.2~\mathrm{nm}$) governing mainly the resonant wavelength value. As mentioned in \ref{append1}, we consider that the numerical aperture value is limited by the illumination objective (N.A.=0.32). The spectral width of the calculated linear transmission spectrum $\overline{T}(\lambda)$ is in good agreement with the experimental results. For the comparison between the experimental and theoretical fluorescence spectra, we had to take into account the characteristics of our confocal microscope. Indeed, it is not possible to reproduce the width and the amplitude of the experimental spectra only using the expression of the overall radiated power $dP_{\mathrm{tot}}/d\lambda$ derived from equation (\ref{radiation}). This is illustrated by the dash theoretical curves of figure \ref{experiment}.(b). We have normalized the theoretical calculations using the maximum of the microcavity spectrum. In this case the theoretical HR mirror emission spectrum, normalized in the same way, does not fit the experimental data. The confocal aperture spatial filtering must be taken into account to obtain a good agreement between the theory and the experimental results \cite{Enderlein00}. The model used to simulate the propagation across the microscope is described in \ref{diffraction}. After fixing the value of $d_0$ obtained from the transmission spectrum fit, the defocusing $\delta z$ (see \ref{diffraction}) is the only adjustable experimental parameter. Note that we have already mentioned that we have experimentally selected the brightest emitters whose the two orthogonal dipoles are parallel to the mirror plane. Consequently we assume in the following calculations that $\Theta=\pi/2$ and $\alpha=0$ as it was the case in the theoretical part. The diameter of the confocal aperture $D=50~\mu\mathrm{m}$ is well adapted for the case of an emitter deposited on the HR mirror. In this case, the pinhole does not truncate the electric field distribution and we obtain an almost perfect collection efficiency for $\delta z=0$. Calculations show that for an emitter inside the cavity, the pinhole diameter is not perfectly adapted. The collection efficiency is thus reduced with respect to the case of the HR mirror. This detrimental effect could be reduced by using a larger confocal aperture. The effect of the LR mirror on the imaging properties of the system is taken into account by using a negative defocusing \cite{Haeberle03}. The theoretical calculations given in figure \ref{experiment}.(b) have been carried out using a defocusing $\delta z=-1500~\mathrm{nm}$ almost corresponding to the optimal value for the overall signal collection. In this case, we obtain a very good agreement simultaneously for the shape of the spectrum and for the relative amplitude between the reference (HR mirror) and the microcavity.

We show that the spectrum of a single NV center can be tailored using a planar microcavity in a similar way than for single molecules~\cite{Chizhik09}. From a qualitative point of view, the fluorescence spectrum is asymmetric and shows profile steepening from its red side which is in good agreement with other experimental observations in microcavity LEDs~\cite{Oulton01}. We also show a good agreement between the experimental results and the theoretical calculations taking into account the optical and geometrical properties of our experimental setup.

We then compared the performances of the sole HR mirror single-photon source, to the one with the emitter inside the microcavity.
\begin{table}
\caption{\label{comparaison}Counting rates of a single NV color center in the case of the two studied configurations: HR mirror alone and microcavity (HR mirror + LR mirror on top). Note that the measurements were done on two different single emitters.} 

\begin{indented}
\lineup
\item[]\begin{tabular}{@{}*{7}{l}}
\br                              
Configuration&Excitation laser&Total photon&Counting rate spectral\cr 
             &power&counting rate&density around $\lambda_0$\cr 
\mr
HR mirror&11.5~mW&140~kcounts/s&0.5~counts/s/nm\cr
Microcavity&10.2~mW&63~kcounts/s&1.0~counts/s/nm\cr 
\br
\end{tabular}
\end{indented}
\end{table}
Table \ref{comparaison} reports the total counting rates (sum of both APDs), and the counting rate spectral density for the two configurations at similar input excitation laser power. We measured 140~kcounts/s for the HR mirror configuration and 63~kcounts/s in the case of the microcavity. We thus observe a reduction of $55\%$ between the HR mirror and the cavity case. The theory gives only a reduction of $27\%$ due to the pinhole effect. The difference might come from the filtering of the parasitic light (e.g. Raman scaterring) in the case of the cavity. Note that we would have obtained an augmentation of $70\%$ without taking into account the pinhole filtering. We however observed an interesting modification of the fluorescence in the cavity configuration: due to the narrowing of the emitted spectrum from $B\approx 75$~nm to $B\approx 22$~nm, the counting rate spectral density is twice larger in the microcavity case, which proves that the emission modes of the nanocrystal are to some extent coupled to some microcavity modes for which the fluorescence rate is enhanced.  

\section{Conclusion}

We have shown theoretically using a simple dipole model embedded in a multilayer system that the fluorescence collection of a single NV diamond color center can be enhanced by a factor of two when it is placed into a moderate finesse ($\mathcal{F}\approx60$) microcavity with respect to its emission when it is located onto a single HR mirror.
The calculations have been done in practical configurations of either Quantum Key Distribution corresponding to a room temperature operation and thus broadband emission of the NV center, or of Quantum Information Processing associated to cryogenic temperatures and monochromatic emission at the zero-phonon line wavelength. 
We also performed room temperature experiments using NV color center in diamond nanocrystals and studied the effect of a planar microcavity on the detected fluorescence light from a single NV center. We have shown that the fluorescence spectrum can be narrowed using such a planar microcavity and  that the counting rate spectral density can be increased compared to the use of a single HR mirror.
Such simultaneous spectrum narrowing and fluorescence spectral density enhancement brought by the microcavity represent a significant increase in the performances of the diamond NV color center based single-photon source in the prospects of open-air QKD applications.

\ack

We thank Jean-Pierre Madrange for the realization of the microcavity mechanical parts, and Thierry Gacoin for the preparation of photoluminescent nanodiamonds sample. We acknowledge helpful discussions with Alexios Beveratos, Rosa Brouri-Tualle, Ga\'etan Messin and Jean-Philippe Poizat.

\appendix 
\section{Linear transmission calculation in the case of a focused illumination}\label{append1}
We give here the expression of the averaged transmission spectrum $\overline{T}(\lambda)$ of the microcavity used to reproduce the experimental results. We used the standard linear transmission $T(\lambda,\theta,P)$ calculated for a plane wave propagating in the $\theta$ direction with a polarization $P=\{\mathrm{TE},\mathrm{TM}\}$ using the transfer matrix method \cite{Yeh77} to evaluate the quantity:
\begin{equation}\label{Transmission_moy}
\overline{T}(\lambda)=\frac{1}{2(1-\cos{\theta_M})}\int_0^{\theta_M}\left[T(\lambda,\theta,\mathrm{TE})+T(\lambda,\theta,\mathrm{TM})\right]\sin\theta d\theta.
\end{equation}
We assume that the beam limitation only comes from the numerical aperture of the microscope objective $\mathrm{O_2}$ used to illuminate the microcavity with the white light as shown in figure \ref{setup} and thus $\sin\theta_M=0.32$.

\section{Theoretical description of the propagation in the imaging system}\label{diffraction}

In this section we give the method used to take into account the effect of the confocal aperture (diameter $D$) and the defocusing on the collection efficiency. The defocusing is taken into account by adding a perturbation to a reference wave surface propagating in our stigmatic optical system as represented in figure \ref{Calcul_diffrac} \cite{Born80, Brokmann_these}. We consider here three wave surfaces: i) $\Sigma_f$ which is the object focal sphere of radius $f$ ii) a plane wave surface $\Sigma$ located in the intermediary space between the microscope objective (focal $f$) and its tube lens of focal $f'$ and iii) the image focal sphere $\Sigma_{f'}$ of radius $f'$.
\begin{figure}[htbp]
\begin{center}
\includegraphics[width=13cm]{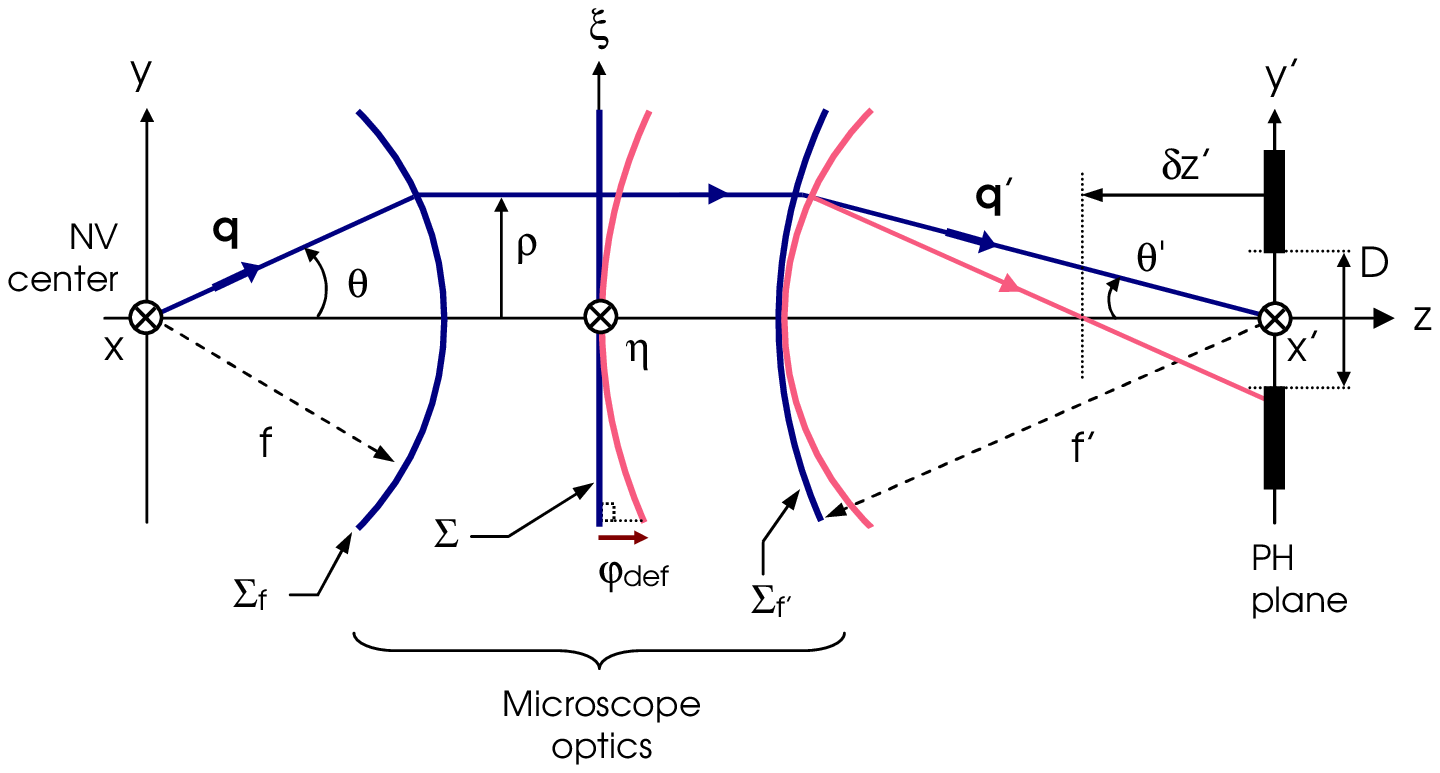}
\caption{The reference focal sphere $\Sigma_f$ is centered on the emitter. This surface is transformed into a plane $\Sigma$ between the objective and the tube lens. In presence of defocusing $\delta z'$, the reference sphere is changed by adding a phase term $\varphi_{\mathrm{def}}$. The confocal aperture (PH) of diameter $D$ is located at the center of the unperturbed reference focal sphere $\Sigma_{f'}$. Both refractive indices of object and image spaces are equal to 1.}\label{Calcul_diffrac}
\end{center}
\end{figure}
We assume a perfect imaging system and these three surface waves are stigmatic. First we calculate the field $\mathbf{E}_f(\mathbf{q})$ (where we note $\mathbf{q}=\mathbf{k}_+/k$) in the $\theta$ direction on the reference sphere $\Sigma_f$ using equations (\ref{sources}). Then we obtain the electric field on the plane wave surface $\Sigma$ \cite{Richards59}:
\begin{equation}
\mathbf{E}_{\Sigma}(\eta,\xi)=\frac{1}{\sqrt{\cos{\theta}}}\mathbf{E}_f(q_x=\eta/f,q_y=\xi/f,q_z)
\end{equation}
where $\eta$ and $\xi$ are the coordinates of the intersection between the emerging ray and $\Sigma$. For a given direction characterized by $\theta'$ and the unitary vector $\mathbf{q}'$, we can write the emerging field on the surface $\Sigma_{f'}$:

\begin{equation}
\mathbf{E}_{f'}(\mathbf{q}')=\sqrt{\cos{\theta'}}\cdot\mathbf{E}_{\Sigma}(\eta=q'_xf',\xi=q'_yf')e^{j\varphi_{\mathrm{def}}(\theta')}.
\end{equation}

The additional phase term $\varphi_{\mathrm{def}}$ accounts for the image space defocusing $\delta z'$:
\begin{equation}
\varphi_{\mathrm{def}}(\theta')=-\frac{2\pi\delta z'}{\lambda}(1-\cos{\theta')}.
\end{equation}
For a given object aperture angle $\theta$, the corresponding image angle $\theta'$ is calculated using: $\rho=f\sin{\theta}=f'\sin{\theta'}$ where $\rho=\sqrt{\eta^2+\xi^2}$ \cite{Richards59}. The defocusing value in the image space $\delta z'$ is connected to the defocusing in the object space $\delta z$ using the sine and Herschel conditions noting that $f'/f=G_t$ where $G_t=40$ is the magnification of the objective $\mathrm{O}_1$. In the image space, the numerical aperture is weak and the paraxial approximation can be used. In particular, the electric field $\mathbf{E}$ in the focal plane $(x',y')$ can be calculated using a Fourier Transform (FT):

\begin{equation}
\mathbf{E_{\mathrm{PH}}}(x'/\lambda,y'/\lambda)=\mathrm{FT}\left[\mathbf{E}_{f'}(\eta/f',\xi/f')\right].
\end{equation}

We can thus obtain the intensity distribution in the image plane by keeping only the $x'$ and $y'$ components of $\mathbf{E}_{\mathrm{PH}}$ \cite{Haeberle03, Born80} which allows the collected spectral density power to be calculated by an integration over the pinhole area:
\begin{eqnarray}
\frac{dP_{\mathrm{tot}}}{d\lambda}(\lambda)&&=\frac{c\epsilon_0}{2}\int\int_{\mathrm{PH}}f(\lambda)\left[\left|E_{\mathrm{PH},x'}(x'/\lambda,y'/\lambda)\right|^2\right.\\
&&+\left.\left|E_{\mathrm{PH},y'}(x'/\lambda,y'/\lambda)\right|^2\right]dx'dy'.\nonumber
\end{eqnarray}

\end{document}